\documentclass[superscriptaddress,preprint,prb]{revtex4}

\usepackage{amsmath}
\usepackage{graphicx}
\usepackage{epsfig}
\usepackage{url}

 \def\bea{\begin{eqnarray}}
 \def\eea{\end{eqnarray}}
\def\beq{\begin{equation}}
\def\eeq{\end{equation}}

 \newcommand{\singlespace}{
 \renewcommand{\baselinestretch}{1}
 \large\normalsize}

\begin{document}
\title{Oblique Collisions of Baseballs and Softballs with a Bat}
\author{Jeffrey R. Kensrud}
\email{jkensrud@gmail.com}
\affiliation{School of Mechanical and
Materials Engineering, Washington State University, Pullman,
Washington 99164}

\author{Alan M. Nathan}
\email{a-nathan@illinois.edu}
\affiliation{Department of Physics, University of Illinois,
Urbana, Illinois 61801}

\author{Lloyd V. Smith}
\email{lvsmith@mme.wsu.edu}
\affiliation{School of Mechanical and
Materials Engineering, Washington State University, Pullman,
Washington 99164}

\date{\today}
\vskip 0.1in

\begin{abstract}
\noindent
Experiments are done by colliding a swinging bat with a stationary baseball or softball.  Each collision was recorded with high-speed cameras, from which the post-impact speed, launch angle, and spin of the ball could be determined.  Initial bat speeds were in the range 63-88 mph, producing launch angles in the range 0$^\circ$-30$^\circ$ and spins in the range 0-3500 rpm.  The results are analyzed in the context of a ball-bat collision model and the parameters of that model are determined.  For both baseballs and softballs, the data are consistent with a mechanism whereby the ball grips the surface of the bat, stretching  the  ball  in  the  transverse  direction and resulting  in  a  spin  that was up to 40\% greater than would be obtained by rolling contact of rigid bodies.  Using a lumped parameter contact model, baseballs are shown to be less compliant tangentially than softballs.  Implications of our results for batted balls in game situations are presented.
\end{abstract}

\maketitle

\section{Introduction}\label{sec:intro}

Scattering experiments have long been a staple for physicists, having been used to probe the structure of objects over a broad range of length and energy scales, from the structure of solids, molecules, and atoms, to the structure of the atomic nucleus, to the structure of the nucleons that make up the nucleus.   Indeed, much of what is known about how the constituents of matter arrange themselves into composite objects have come from scattering experiments.

Students are often surprised to learn that some of the same principles that apply to subatomic collisions also apply to collisions of macroscopic objects.  Some of these principles are obvious, such as conservation of momentum and angular momentum in both elastic and inelastic collision.  Others are not as obvious.  One question that often comes up regarding inelastic collisions is this:  Where did the missing kinetic energy go?  In subatomic collisions, the missing energy often goes to exciting internal degrees of freedom of one or more of the colliding objects.  It is interesting that the same is often true in the collision of macroscopic objects.  For example, a ball is dropped onto the floor and it bounces to a fraction of its initial height.  The missing energy went into exciting the molecules that make up the ball, eventually appearing as heat.  The transfer of energy from external to internal degrees of freedom is a theme that shows up in diverse areas of physics.

Not only is the study of the internal degrees of freedom interesting from a purely physics point of view, but it also often has practical value.  This is especially true in the collisions of sports balls with striking objects, such as rackets, clubs, or bats, where it is often important to maximize the speed of the struck ball by minimizing the loss of kinetic energy.  It is also often important to maximize the spin of a struck ball in an oblique collision with a surface.  Minimizing the loss of kinetic energy and maximizing the post-impact spin means maximizing both the normal and tangential coefficients of restitutions, a topic we will explore in detail in this article for the collision of a baseball or softball with a bat.

There have been very few investigations of oblique collision between a ball and bat.  Both Sawicki, {\it et al.}\cite{Sa03} and Watts and Baroni\cite{Watts} did theoretical calculations of oblique collisions, using a model  in which the ball and bat mutually roll as the ball leaves the bat.  Subsequent experiments by Cross and Nathan\cite{CrossNathan06} at very low speed and by Nathan, {\it et al.}\cite{Na12} at much higher speed have shown that balls do not roll as they bounce but instead grip the surface, resulting in an enhancement to the post-impact spin.  In both these experiments, a ball was scattered from a bat which was initially at rest.  We investigate this topic in a new experiment, in which a moving bat impacts a ball that is initially at rest.  The results confirm the previous experiments that show an enhancement to the spin.  A preliminary version of this work has appeared elsewhere.\cite{Ke13}

\section{Experimental Method and Data Reduction}\label{sec:expt}

\begin{figure}[h]
\vspace{0.4in}
\singlespace
\includegraphics[angle=0,width=0.8\textwidth]{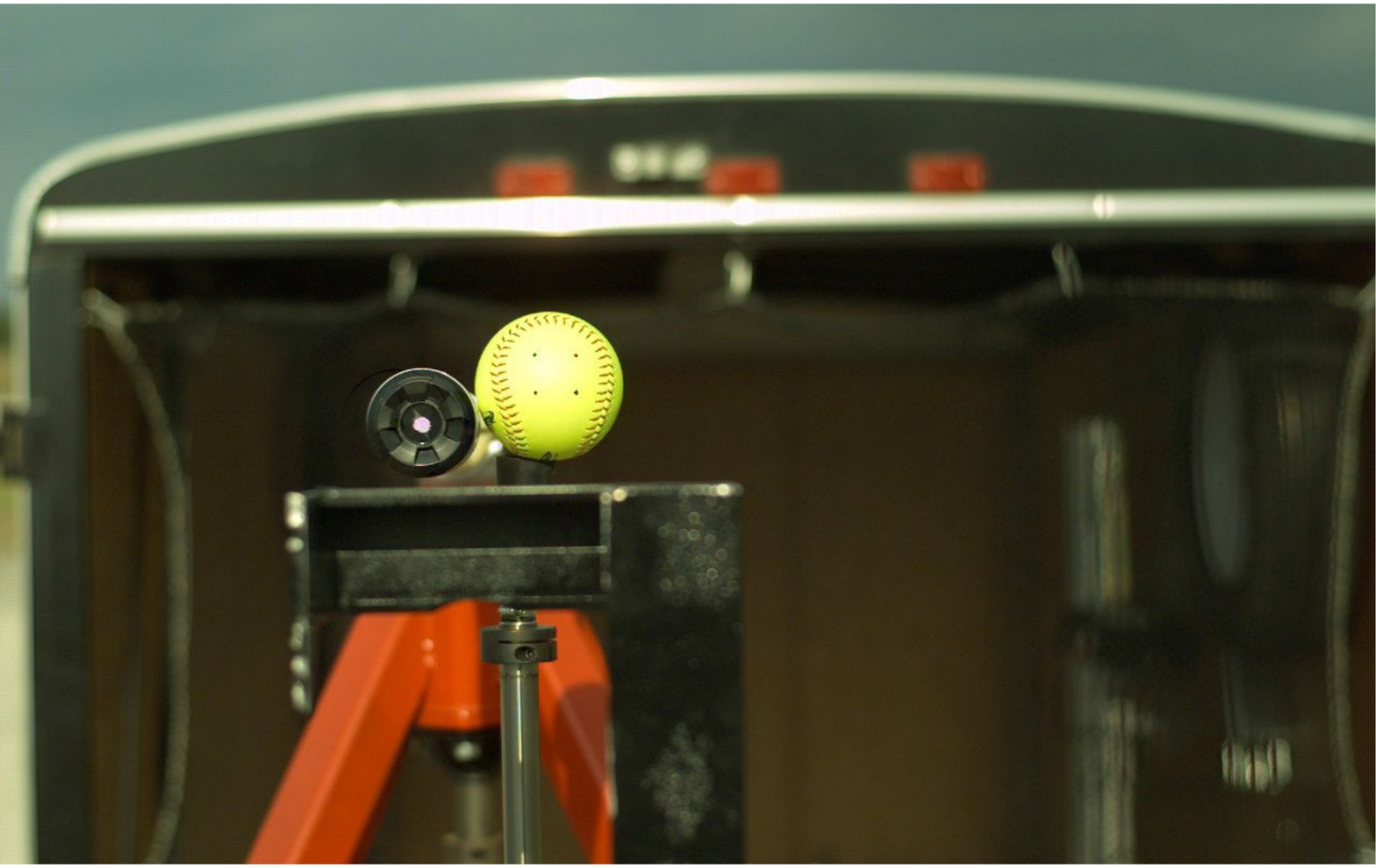}
\includegraphics[angle=0,width=0.8\textwidth]{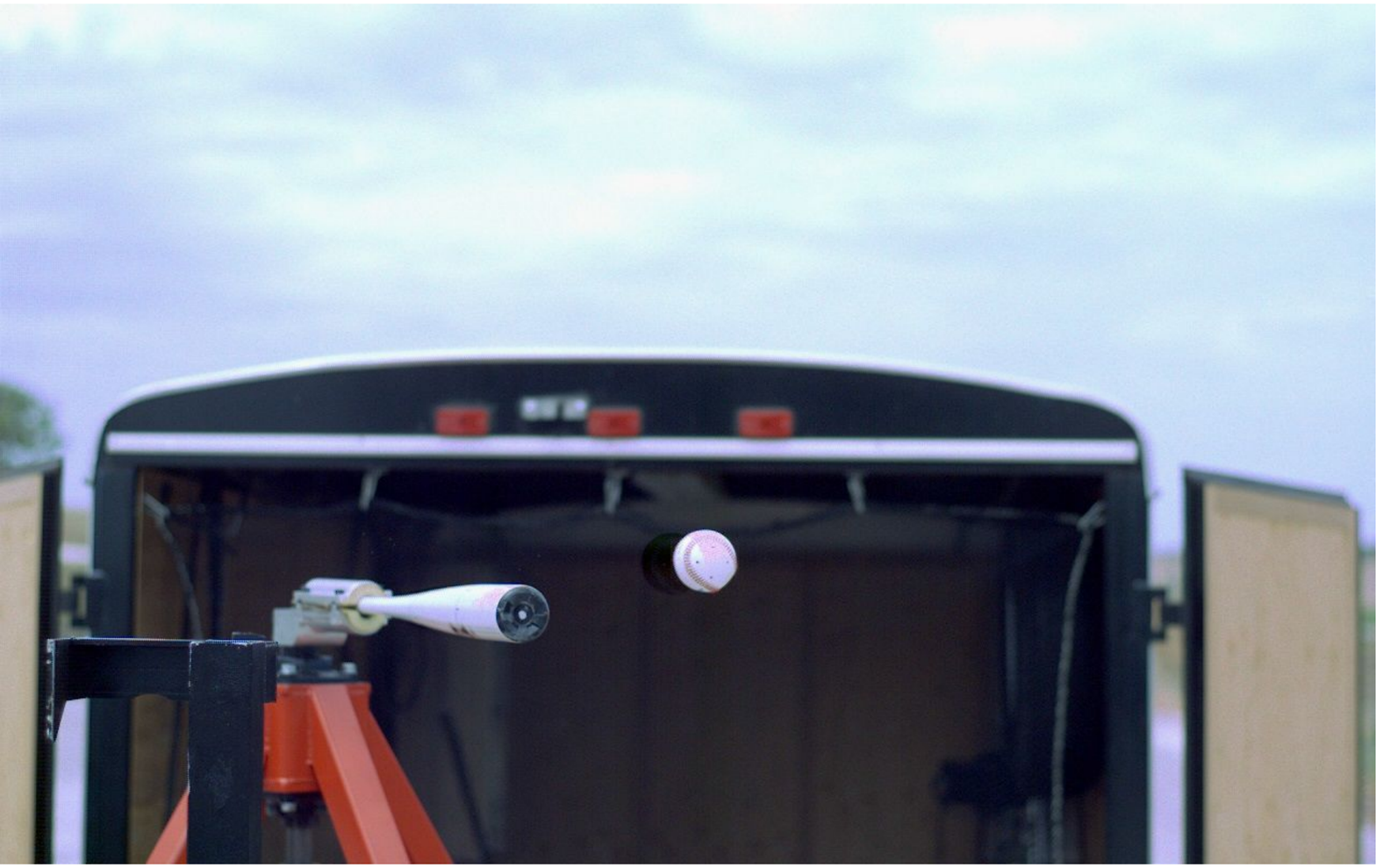}
\caption{\label{fig:cameras}Representative views from Camera 1 (top) and Camera 2 (bottom) used to measure bat and ball speed and rotation.}
%\end{center}
\end{figure}

The aim of this work was to measure the velocity and spin rate of the batted ball at collision speeds representative of play. A machine was used to swing bats against a stationary ball resting on a tee. Bats were swung to achieve speeds between 63 and 88 mph (28 and 39 m/s, respectively) at the impact location. Bats were fastened onto a rotating pivot with a flexible clamp. A 10 mm thick piece of rubber was placed between the clamp and the bat handle, keeping the bat from slipping out of the machine and allowing compliancy to simulate a batter’s hands. Each impact was measured with two cameras (1200 $x$ 800 pixels) placed in the plane of the swinging bat.  Camera 1 was set to a viewing plane of 23 by 35 inches (0.58 by 0.89 m) and recorded the bat-ball impact at 3000 frames per second. It was aligned collinearly with the bat at impact, as shown in Fig.~\ref{fig:cameras}. Camera 2, also shown in Fig.~\ref{fig:cameras}, was set to a viewing plane of 39 by 59 inches (1.0 m by 1.5 m) and recorded the impact at 1000 frames per second.

Each batted-ball event was tracked with the ProAnalyst 2-D tracking software. The bat tip speed and swing plane angle were determined from Camera 1 using a tracking dot on the end cap of the bat.  Since the impact location along the bat and the location of the bat rotation axis are known, the speed of the bat $v_{bat}$ at the impact location could be determined.  The hit ball speed $v_{ball}$ and the angle of the ball with the horizontal were also obtained from Camera 1.  From this information, the angle $\theta$ of the outgoing ball with respect to the swing direction could be determined, as shown in Fig.~\ref{fig:geom}. Camera 2 was used to measure the spin rate $\omega$ of the ball in the following manner.  Each ball was marked with a series of four tracking dots forming the outside corners of a 1.5-inch (38-mm) square pattern. The ball was placed on the tee so that the center of rotation of the ball after impact was within the interior of the square pattern defined by the four dots.  Two of the four dots were tracked for each hit. Impacts where at least one dot did not rotate about the center of rotation were not included in the data set. Each dot's coordinates were used to calculate the angle of rotation from each video frame which were then used to find $\omega$. The spin was averaged over the time the ball left the bat to the time when the ball left the viewing plane (typically 15 frames).

The properties of the balls and bats used in the study are presented in Tables I and II, respectively.  Both baseball and softballs were used.  Three different baseball and one softball bat were utilized.  The baseball bats included a wood bat, a metal bat with a smooth surface, and a metal bat with a rough surface.  The softball bat was made of a composite material and had a smooth surface.  Standard methods\cite{bifilar} were used to measure the moment of inertia of each bat about the central axis ($I_z$) and about an axis perpendicular to the central axis and passing through the center of mass ($I_0$).

\begin{figure}[h]
%\begin{center}
\vspace{0.4in}
\singlespace
\includegraphics[angle=0,width=0.7\textwidth]{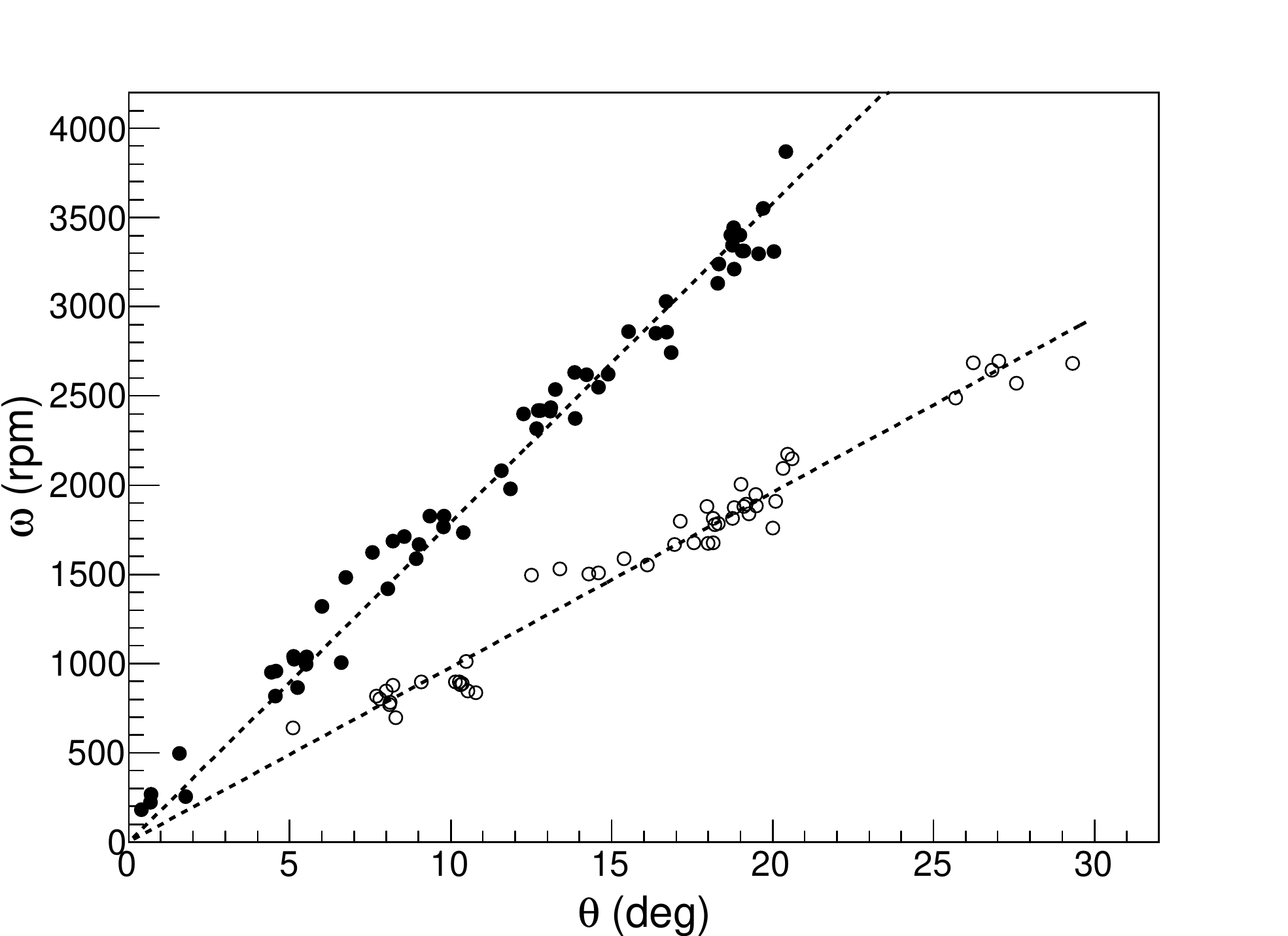}
\caption{\label{fig:omega-theta}Normalized spin of the batted ball as a function of launch angle.  The closed and open points are for baseballs and softballs, respectively.  The dashed lines are linear fits with slopes 179 and 98 rpm/deg.}
%\end{center}
\end{figure}

\begin{figure}[h]
%\begin{center}
\vspace{0.4in}
\singlespace
\includegraphics[angle=0,width=0.7\textwidth]{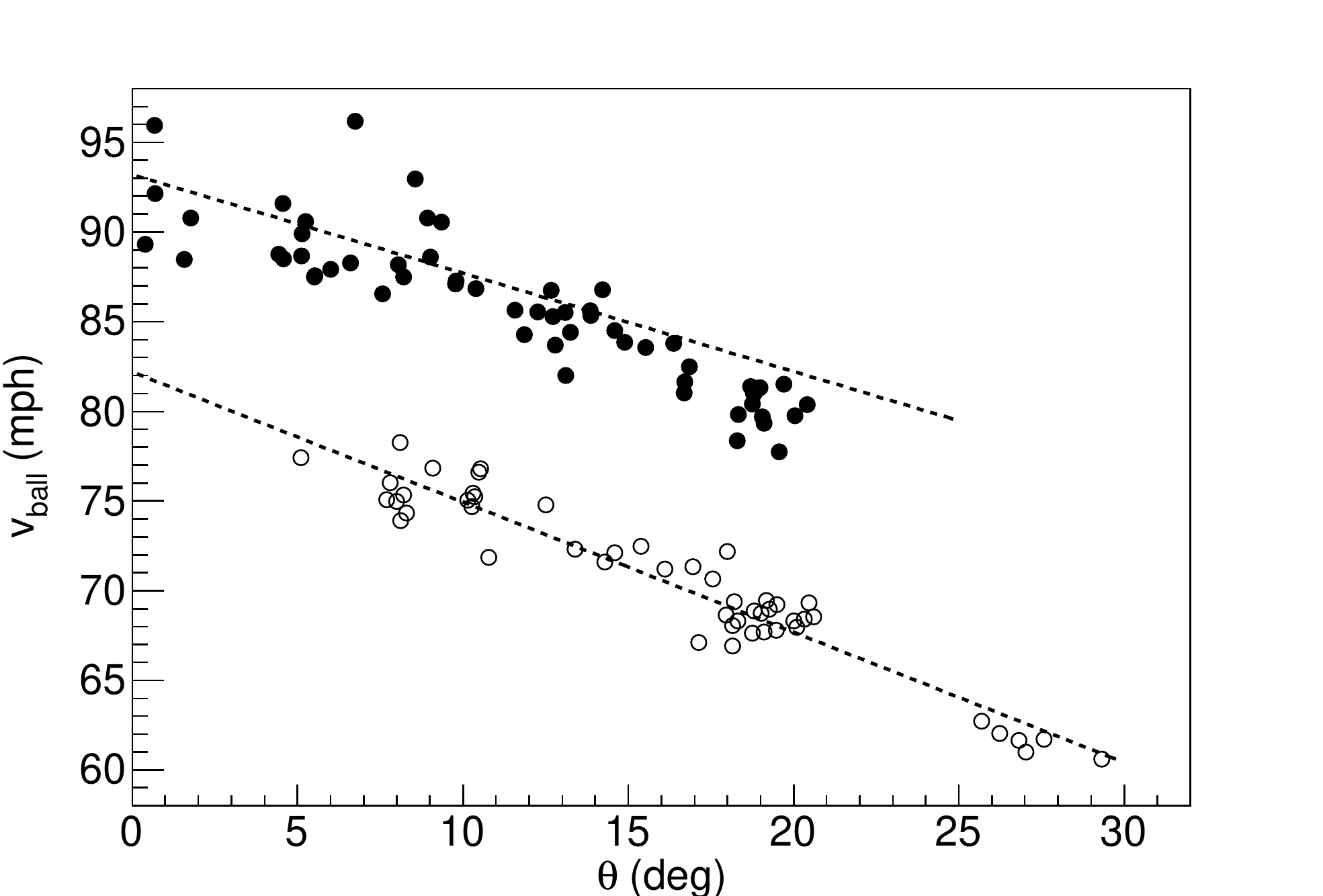}
\caption{\label{fig:vf-theta}Normalized exit speed of the batted ball as a function of launch angle.  The closed and open points are for baseballs and softballs, respectively.  The dashed lines are linear fits, with slopes -0.65 and -0.72 mph/deg}
%\end{center}
\end{figure}

The reduced data for the experiment are shown in Figs.~\ref{fig:omega-theta}-\ref{fig:vf-theta}, which show, respectively, the spin rate and exit speed of the ball as a function of the scattering angle $\theta$.  As will be shown in the next section, both the spin rate and the exit speed scale linearly with the initial bat speed.  Since a range of speeds were used in the experiment, the spin rates and exit speeds in Figs.~\ref{fig:omega-theta}-\ref{fig:vf-theta} have been normalized to a bat speed of 77 mph (34 m/s) by multiplying the actual spins and speed by $\left [v_{bat}/77\mbox{ mph}\right ]$.   For both baseballs and softballs, the normalized spins are remarkably linear over the full range of $\theta$, a topic that will be discussed in Sec.~\ref{sec:results}.

\section{Ball-Bat Collision Formalism}\label{sec:model}
\subsection{Kinematics}

\begin{figure}[h]
%\begin{center}
\vspace{0.4in}
\singlespace
\includegraphics[angle=0,width=0.7\textwidth]{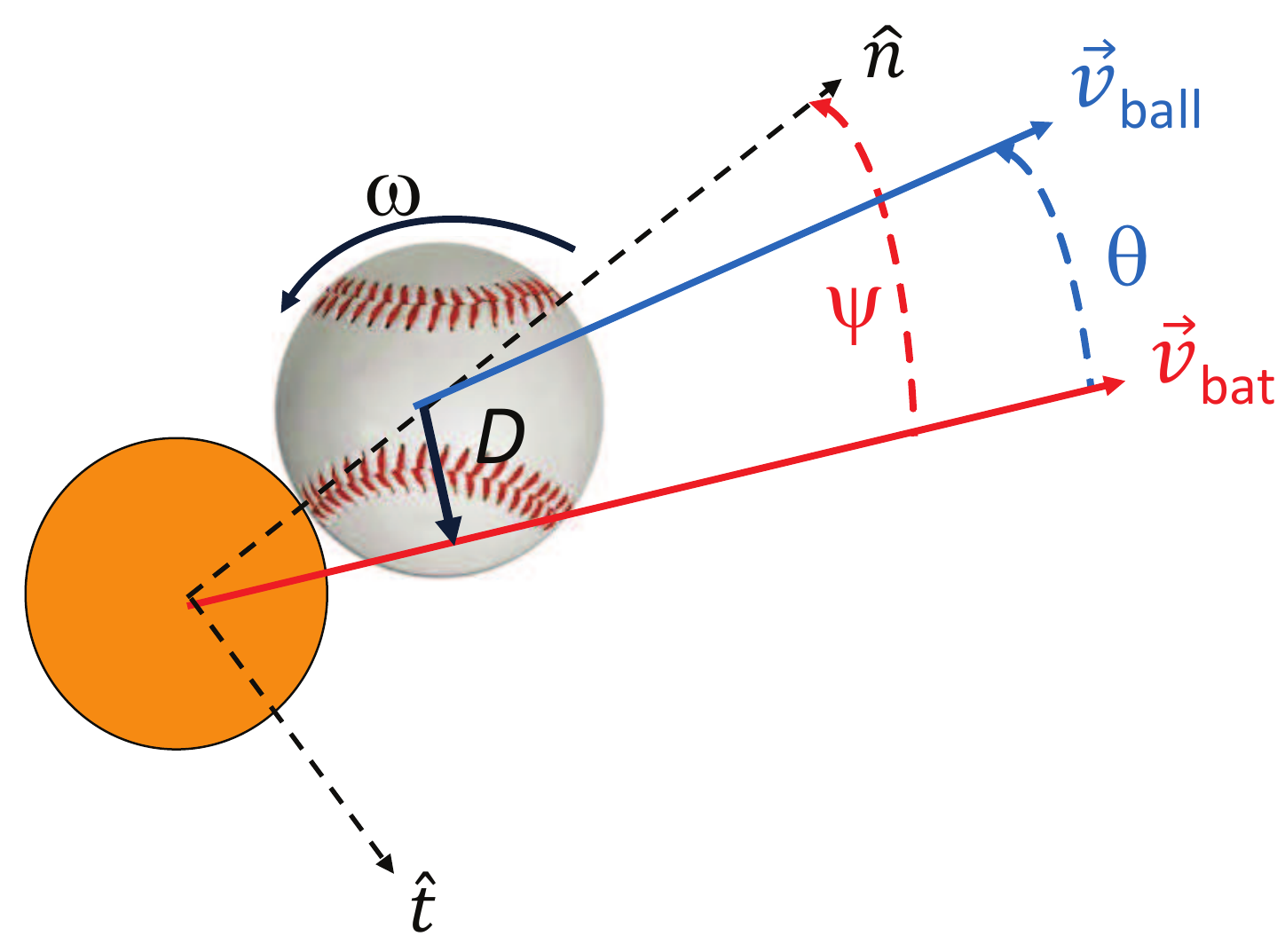}
\caption{Geometry for the ball-bat collision experiment, showing the initial velocity of the bat $\vec{v}_{bat}$, the velocity of the struck ball $\vec{v}_{ball}$, the scattering angle $\theta$, and the post-impact spin $\omega$.  Also shown are the unit vectors $\hat{n}$ and $\hat{t}$ normal and transverse, respectively, to the bat surface, and the angle $\psi$ between the bat direction and the normal.  The parameter D is the perpendicular distance between the center of the ball and $\vec{v}_{bat}$}
%\end{center}
\label{fig:geom}
\end{figure}

The goal of the analysis is to interpret the experimental results in the context of the collision formalism described in detail by Cross and Nathan.\cite{CrossNathan06}  The collision geometry is shown in Fig.~\ref{fig:geom}.  The velocities of the ball and bat, $\vec{v}_{ball}$ and $\vec{v}_{bat}$, are decomposed into components along the normal axis $\hat{n}$, defined as the line connecting the centers of the bat and ball at impact, and the transverse axis $\hat{t}$, which is in the scattering plane and orthogonal to $\hat{n}$.  Neither the $\hat{n}$ nor the $\hat{t}$ axes are directly measured.  However, they can be determined from the conservation of angular momentum of the ball about the contact point.  Assuming the ball-bat interaction indeed acts at a point and the center of mass is at the geometric center of the ball, then the post-collision angular momentum of the ball, which is initially at rest with zero angular momentum, must vanish.  That is

\beq
I_{ball}\omega\,=\,mv_{ball,t}r\,=\,mv_{ball}r\sin(\psi-\theta) \, ,
\label{eq:angmom}
\eeq
where $r$ and $m$ are the radius and mass of the ball, respectively, and $I_{ball}$ is the moment of inertia of the ball about its center:

\beq
I_{ball}\,=\,\alpha mr^2 \,
\label{eq:iball}
\eeq
and $\alpha$=0.4, the value for a uniform sphere.\cite{Brody05}  Since $v_{ball}$, $\theta$, and $\omega$ are all measured quantities, Eq.~\ref{eq:angmom} can be solved for $\psi$, the angle of $\vec{v}_{bat}$  with respect to the $\hat{n}$ axis.  Note that the angular momentum due to the spin points out of the plane of the diagram and is canceled by the angular momentum due to the center of mass motion, which points in the opposite direction.  The angle $\psi$  is related to the ball-bat offset $D$ by\cite{D}
\beq
D\,=\,(r+R)\sin\psi \, .
\label{eq:D}
\eeq
The maximum values of $\psi$ and $D$ in the present experiment are given in Table I.

The normal part of the collision utilizes the expression\cite{CrossNathan06}
\beq
v_{ball,n}\,=\,\left [\frac{1+e_y}{1+r_y}\right ]v_{bat,n} \, ,
\label{eq:vballn}
\eeq
where $e_y$ is the normal coefficient of restitution and $r_y$ is a kinematic factor associated with the recoil of the bat in the normal direction and is given by
\beq
r_y\, = \, m\Big ( \frac{1}{M} +
\frac{b^2}{I_0}\Big ) \, .
\label{eq:ry}
\eeq
Analogously, the transverse part of the collision utilizes the expression\cite{CrossNathan06}
\beq
v_{ball,t}+r\omega\,=\,\left [\frac{1+e_x}{1+r_x}\right ]v_{bat,t} \, ,
\label{eq:vballt}
\eeq
where $e_x$ is the transverse coefficient of restitution and $r_x$ is a kinematic factor associated with the recoil of the bat in the transverse direction and is given by
\beq
r_x\, = \, \frac{m\alpha}{1+\alpha}\Big ( \frac{1}{M} +
\frac{b^2}{I_0} + \frac{R^2}{I_z}\Big ) \, .
\label{eq:rx}
\eeq
The sign convention is such that all the velocities and spin in Eq.~\ref{eq:vballt} are positive for the case shown in Fig.~\ref{fig:geom}.
The mass $M$, radius $R$, moments of inertia $I_0$ and $I_z$, and impact location $b$ for each of the four bats are given in Table II.
Eqs.~\ref{eq:angmom} and \ref{eq:vballt} can be combined to arrive at the following equations:
\beq
\omega\,=\,(1+e_x)\left [\frac{v_{bat}\sin\psi}{r(1+r_x)(1+\alpha)}\right ] \, ,
\label{eq:omega}
\eeq
and
\beq
v_{ball,t}\,=\,(1+e_x)\left [\frac{\alpha v_{bat}\sin\psi}{(1+r_x)(1+\alpha)}\right ] \, .
\label{eq:vxf}
\eeq
Alternately, the two preceding equations can be combined to obtain
\beq
\frac{v_{ball,t}+r\omega}{v_{bat,n}} \, = \, (1+e_x)\left [\frac{\tan\psi}{1+r_x} \right ] \, .
\label{eq:vtf}
\eeq

\subsection{Slipping, Rolling, and Gripping}

 Other than purely kinematic terms, the collision is determined by the coefficients of restitution, $e_y$ and $e_x$, which govern energy conservation in the normal and transverse directions.\cite{Cross15}  Purely normal collisions (those with $D=\psi=0$), have no transverse velocity component and have been studied extensively in the literature, providing considerable information about $e_y$ for collisions between a baseball or softball with a bat.\cite{Nathan03,Smith09}  By contrast, there have been very few studies of oblique collision.  Accordingly, the present physics analysis will focus on the transverse part and particularly $e_x$, which
plays an important role in determining the spin of the struck ball. Therefore it is useful at this stage to discuss its physical significance.

It is defined as the negative of the ratio of relative tangential velocity at the contact point after the collision to that before the collision.\cite{Cr02a}   When the bat makes contact with the stationary ball, it exerts a force $N$ on the ball in the $\hat{n}$ direction.  Suppose the bat has nonzero tangential velocity $v_{bat}\sin\psi$, and consider the case where the ball and bat are completely rigid in the transverse direction.  Under such circumstances, the ball and bat will initially slide along their mutual surfaces.  As a result, the ball exerts a frictional force $F= \mu_kN$ on the bat in the -$\hat{t}$ direction, and the bat exerts an identical force $F$ on the ball in the +$\hat{t}$ direction, causing it to accelerate in that direction and to spin as shown in Fig.~\ref{fig:geom}.   Here $\mu_k$ is the coefficient of sliding friction.  If $F$ brings the sliding to a halt prior to the ball leaving the surface, then $F$ drops to zero and the ball rolls along the surface until it leaves the bat.  In that case, the transverse impulse is related to the normal impulse by $\int Fdt < \mu_k \int Ndt$.  Moreover, since the relative tangential velocity is zero when rolling, then $e_x=0$.  If $F$ is insufficient to bring the sliding to a halt before the ball leaves the bat, a condition referred to as ``gross slip'', then the final and initial relative tangential velocities have the same sign so that $e_x<0$ and the spin is reduced relative to the rolling case.  In that case, $\int Fdt = \mu_k \int Ndt$.  The gross slip and rolling cases are the only possibilities for a rigid ball and were the only ones considered by Watts\cite{Watts} and by Sawicki, {\it et al.}\cite{Sa03}

For a ball with tangential compliance, a third case is possible in which the relative tangential motion drops suddenly to zero as the ball grips the surface, either initially or after a period of sliding, as kinetic energy associated with the transverse motion is converted to potential energy associated with the tangential stretching of the ball.\cite{Cr02a}  Under such circumstances, the contact point is at rest, held in place by a force $F$ due to static friction, so that $F\le \mu_sN$, where $\mu_s$ is the coefficient of static friction.  To analyze such a situation in detail requires a dynamic model,\cite{Ma81,Stronge00,Stronge94} an example of which will be discussed in Sec.~\ref{sec:discussion}. Depending on the details, the resulting $e_x$ can be positive, so that the final spin is enhanced relative to the rolling case, a condition referred to as ``overspin''.   For all three scenarios (rolling, gross slip, and gripping), $\int Fdt \le \mu \int Ndt$, where the distinction between the coefficients of sliding and static friction has been ignored and where equality is achieved only in the special case of gross slip.  A positive $e_x$ corresponds to a reversal of the initial relative tangential velocity, which can only occur if there is tangential compliance.

\section{Results}\label{sec:results}

\begin{figure}[h]
%\begin{center}
\vspace{0.4in}
\singlespace
\includegraphics[angle=0,width=0.7\textwidth]{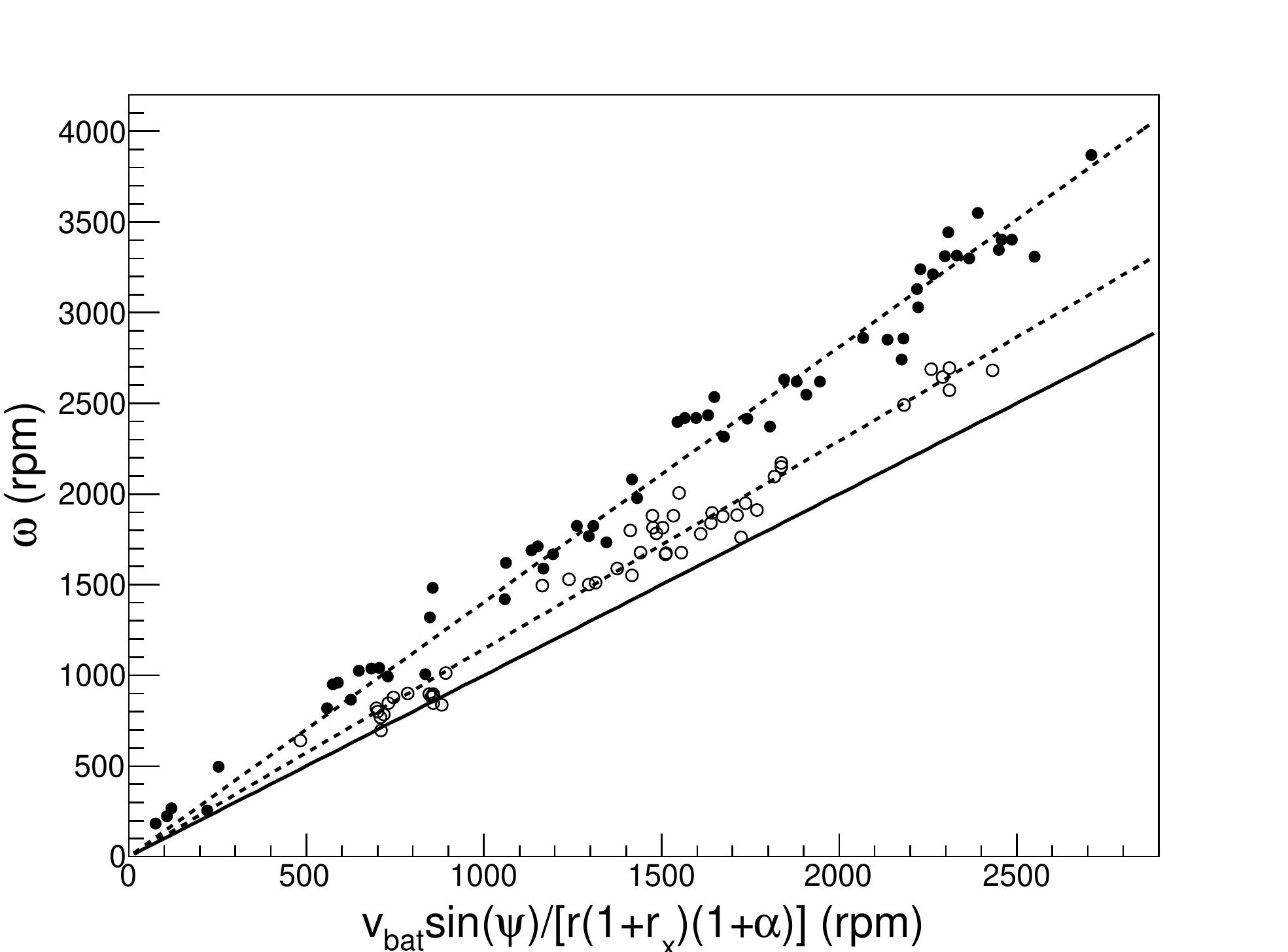}
\caption{\label{fig:omegavbatT}Spin of the batted ball as a function of the term in brackets on the RHS of Eq.~\ref{eq:omega}.  The closed and open points are for baseballs and softballs, respectively.  The slopes of the dashed lines, which are equal to 1+$e_x$, are 1.405 and 1.146 for baseballs and softballs, respectively. For reference, the solid line has unit slope, which would be expected if $e_x$ were zero.}
%\end{center}
\end{figure}

The physically interesting quantity to be derived from these data, $e_x$, is examined in Fig.~\ref{fig:omegavbatT}, where the spin is plotted against the RHS of Eq.~\ref{eq:omega} for baseballs and softballs separately but averaged over all bats.  The dashed curve is a linear fit, where Eq.~\ref{eq:omega} shows that the slope equals 1+$e_x$, from which we find

\bea
\langle e_x \rangle \,& =& \, 0.405\pm 0.010 \mbox{  (baseballs)}  \nonumber\\
&=&\, 0.146\pm 0.009 \mbox{  (softballs)}\, ,
\label{eq:ex}
\eea
where the average is taken over all bats.  Linear fits were also done to individual bats, and the results are given in Table~II.  One simple way to interpret these numbers is that the spin of the batted ball is enhanced relative to the ``sliding-then-rolling'' scenario (i.e., $e_x=0$) by 40\% for baseballs and 15\% for softballs.  For baseballs, this value of $e_x$ exceeds that obtained at comparable speeds.\cite{Na12}  To our knowledge, this is the first such measurement of $e_x$ for softballs.

\begin{figure}[h]
%\begin{center}
\vspace{0.4in}
\singlespace
\includegraphics[angle=0,width=0.7\textwidth]{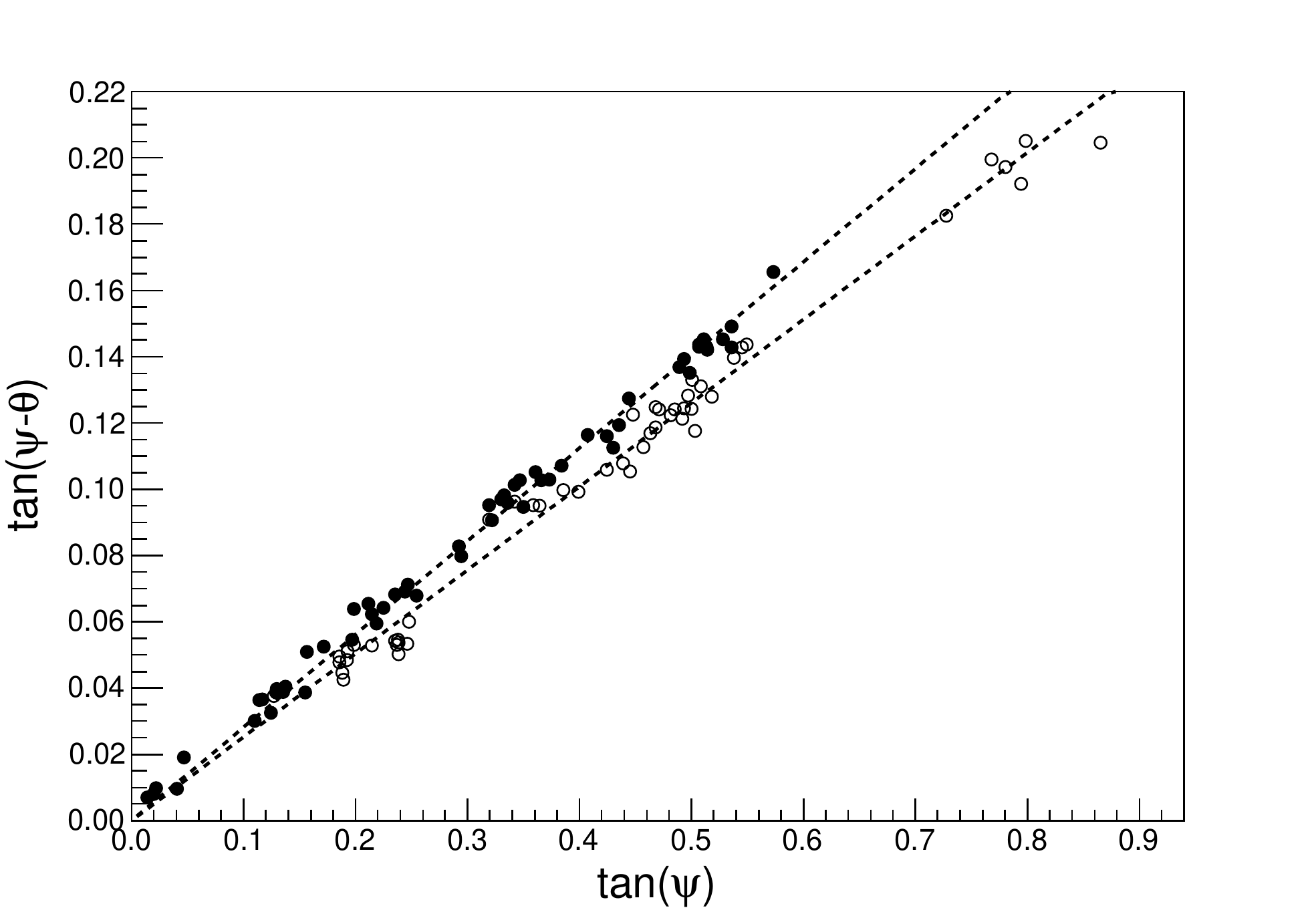}
\caption{\label{fig:thetapsi}Plot of $\tan(\psi-\theta)$ vs. $\tan(\psi)$.  The closed and open points are for baseballs and softballs, respectively.  The dashed lines are linear fits with slopes 0.281 and 0.252 for baseballs and softballs, respectively.  The slopes are proportional to the ratio (1+$e_x$)/(1+$e_y$), as indicated in Eq.~\ref{eq:ratio}.}
%\end{center}
\end{figure}

Given that the data are consistent with $e_x\ge 0$, a gross slip scattering mechanism can be ruled out.  From the discussion in Sec.~\ref{sec:model}, this result can be used to establish a lower limit on the size of $\mu_k$ equal to the ratio of transverse to normal impulses imparted to the ball:

\beq
\mu_k \, \ge \, \frac{\int{Fdt}}{\int{Ndt}}\,=\, \frac{v_{ball,t}}{v_{ball,n}} \,=\, \tan(\psi-\theta)\, ,
\label{eq:muk}
\eeq
where equality is achieved for gross slip.  On the other hand, the RHS of Eq.~\ref{eq:muk} can be expressed as follows, using Eq.~\ref{eq:vballn} and \ref{eq:vxf}:
\beq
\mu_k \, \ge \, \tan(\psi-\theta) \,=\, \left [\left (\frac{\alpha}{1+\alpha}\right )\left (\frac{1+e_x}{1+r_x}\right )\left (\frac{1+r_y}{1+e_y}\right )\right ]\tan(\psi) \, .
\label{eq:ratio}
\eeq
A plot of $\tan(\psi-\theta)$ versus $\tan(\psi)$ is shown in Fig.~\ref{fig:thetapsi}, from which a lower limit of 0.15 and 0.20 can be placed on $\mu_k$ for baseball and softballs, respectively.  This limit is not particularly interesting, since $\mu_k$ is likely significantly larger.\cite{CrossNathan06}  Nevertheless, it does show that considerably larger impact angles $\psi$, and correspondingly larger post-impact spin rates $\omega$ and launch angles $\theta$, can be achieved before gross slip sets in.  For example, if $\mu_k$ were 0.50,\cite{CrossNathan06} the corresponding maximum $\psi$ prior to gross slip would be about 60$^\circ$, leading to $\omega=6400$ rpm and $\theta=34^\circ$.

As noted earlier, $\omega$ is remarkably linear in $\theta$ over the full range of measurements in Fig.~\ref{fig:omega-theta}.  In the small-angle approximation, this result follows immediately from Eqs.~\ref{eq:omega} and \ref{eq:ratio}, which taken together imply $\omega \propto \psi \propto \theta$.  A further observation is that for a given type of ball (baseball or softball), $\omega$ depends only weakly on the bat, despite $I_z$ values differing by a factor of 2 (see Table II).  Keeping in mind that $I_z$ enters the formalism only through the recoil factor $r_x$ (Eq.~\ref{eq:rx}) and that $\omega$ depends only on 1+$r_x$ (Eq.~\ref{eq:omega}), a large sensitivity of $\omega$ to $I_z$ is not expected.

\section{Discussion}
\label{sec:discussion}
\subsection{A lumped parameter model}
In this section we present the results of the dynamic model of Stronge,\cite{Stronge00,Stronge94} in a geometry in which a spherical ball is incident on a massive rigid surface.   For the actual calculations presented below, the formalism has been modified appropriately to accomodate the present experiment, where a bat is incident on a stationary ball.  Nevertheless, for ease of presentation the discussion will be in the context of Stronge's original geometry.

\begin{figure}[h]
%\begin{center}
\vspace{0.4in}
\singlespace
\includegraphics[angle=0,width=0.55\textwidth]{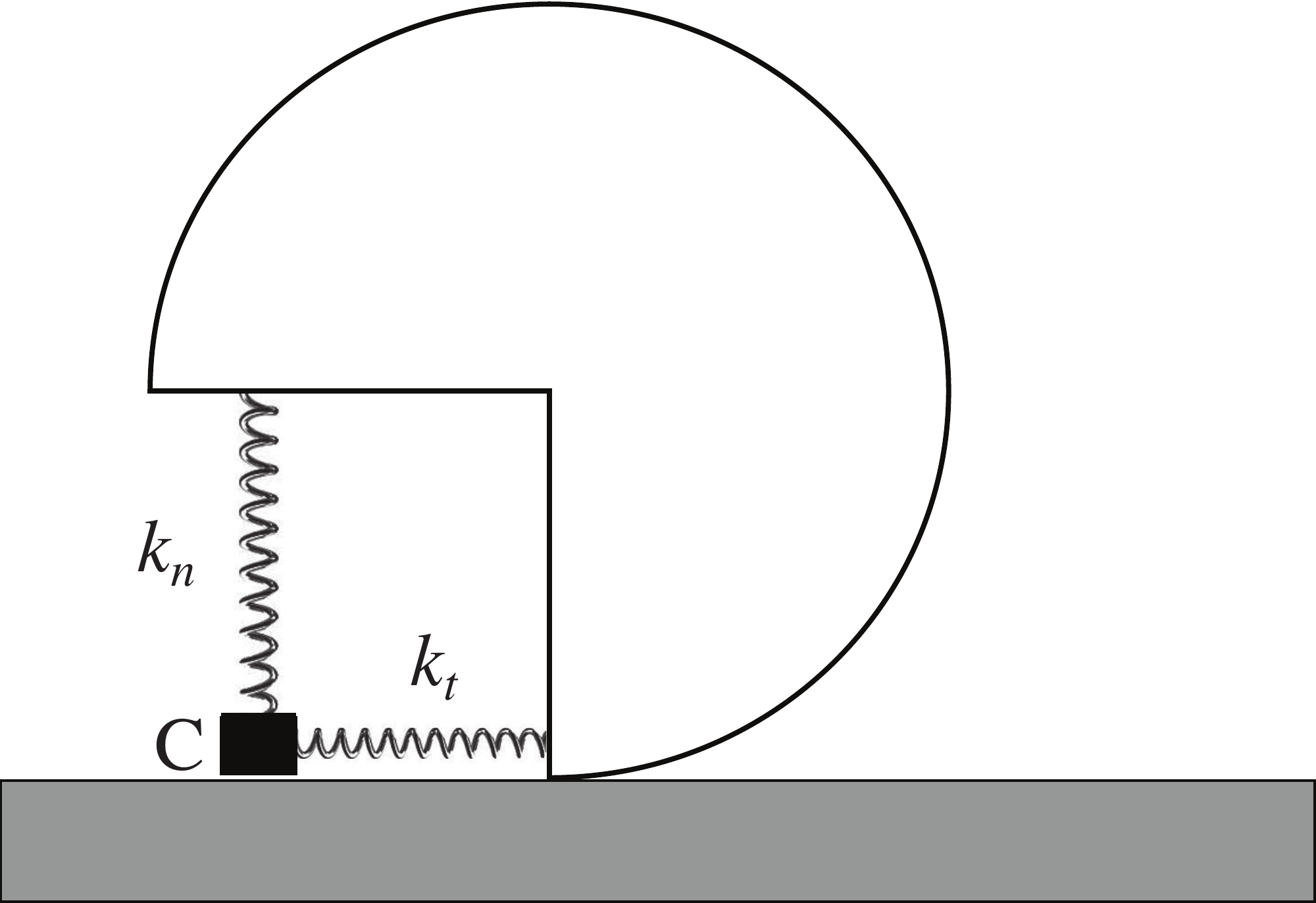}
\caption{\label{fig:Stronge}Geometry for the lumped parameter model of Stronge.  The ball is coupled to a massless contact point C (denoted by the black box) by two springs with force constants $k_n$ and $k_t$.  The contact point either slips on or grips the rough massive surface, denoted by the cross-hatched area.}
%\end{center}
\end{figure}

The collision is treated as a lumped parameter model, where a sphere ($\alpha=0.4$) of mass $m$ is incident at angle $\psi$ on a flat surface.  The sphere is coupled to a massless contact point $C$ via two linear springs, one each for the normal and tangential components of the collision, with force constants $k_n$ and $k_t$, respectively, as depicted in Fig.~\ref{fig:Stronge}.    The normal spring has dissipation, resulting in the normal coefficient of restitution $e_y$, whereas the tangential spring is perfectly elastic.  Besides $e_y$, the other essential parameters of the model are the coefficient of friction $\mu$ and the ratio of normal to transverse spring constants $\eta^2=k_n/k_t$.  The natural vibration frequencies are $\Omega_n=\sqrt{k_n/m}$ and $\Omega_t=\Omega_n\sqrt{3.5/\eta^2}$.

Upon initial contact, the normal spring compresses, then recovers to its relaxed length, terminating the collision after a total contact time $t_f=t_c(1+e_y)$, where $t_c=(\pi/2)/\Omega_n$ is the time to maximum compression.  Since $C$ is massless, there can be no net force on it.  Therefore the transverse force due to the stretching or compressing of the tangential spring is balanced by an opposite frictional force, either static friction if $C$ is at rest (``stick'') or sliding friction if $C$ is moving (``slip'').

There are three distinct cases considered by Stronge.  First when $\tan\psi<\mu\eta^2$, the initial state is stick, which persists until the static friction needed to fix $C$ to the surface exceeds its limit, whereupon $C$ enters the slip state for the remainder of the collision (``stick-slip'').  Second when $\mu\eta^2\le\tan\psi<3.5\mu (1+e_y)$, the initial state is slip, followed later by a stick state, then followed by the slip state for the remainder of the collision (``slip-stick-slip'').  Third, when $\tan\psi\ge 3.5\mu (1+e_y)$ the initial slip state persists throughout the entire collision (``gross slip'').  The conditions for the transition between stick and slip states is determined by $\mu$ and by the relative sizes of the forces from the two springs.

The motion in both the normal and transverse directions can be calculated, either analytically\cite{Stronge94} or numerically, the latter being the approach used here.  For given input parameters and incident angle $\psi$, the equations of motion are integrated until the surfaces separate, obtaining the final transverse speed.  For small $\psi$, the final speed is insensitive to the size of $\mu$, as long as the initial stick criterion is satisfied.  The most crucial parameter is $\eta^2$, which essentially determines the phase of the transverse vibration at separation.

An interesting extreme example is that of a superball, for which $\eta^2=3.5$ and $e_y$=1, implying $\Omega_t=\Omega_n$ and no energy loss for the normal part of the collision.  Therefore, assuming the initial stick criterion is satisfied, at the moment of separation the transverse spring is exactly at its equilibrium position, with the mass moving with the same speed as initially but in the opposite direction.  In the language of Sec.~\ref{sec:model}, $e_x=1$ and the initial kinetic energy is completely conserved, both in the normal and transverse directions.\cite{Ga69}  Cross gives an excellent discussion of this effect in a recent article.\cite{Cross15}

\begin{figure}[h]
    \centering
    \begin{minipage}{.5\textwidth}
        \centering
        \includegraphics[width=\textwidth]{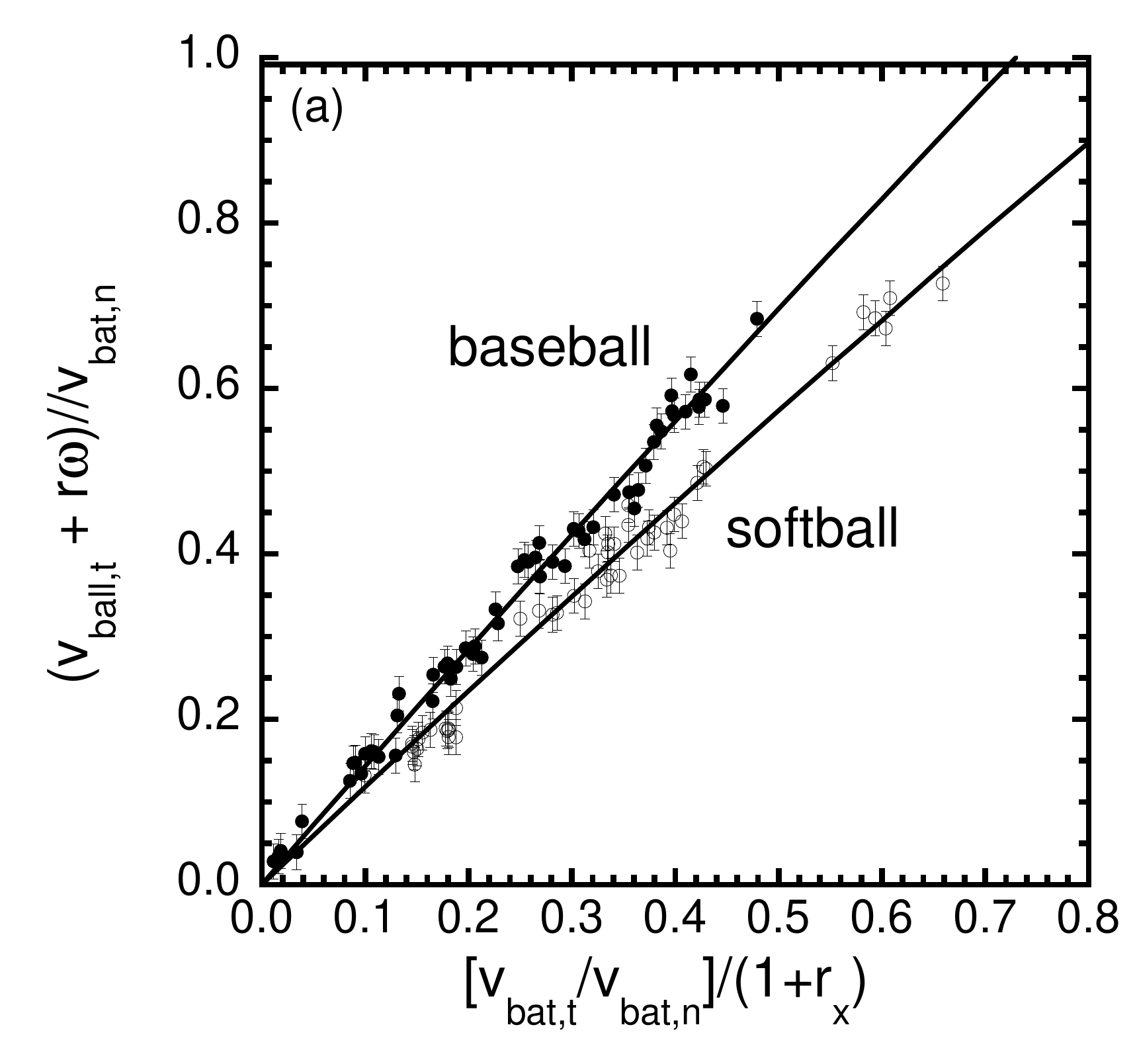}
    \end{minipage}%
    \begin{minipage}{0.5\textwidth}
        \centering
        \includegraphics[width=\textwidth]{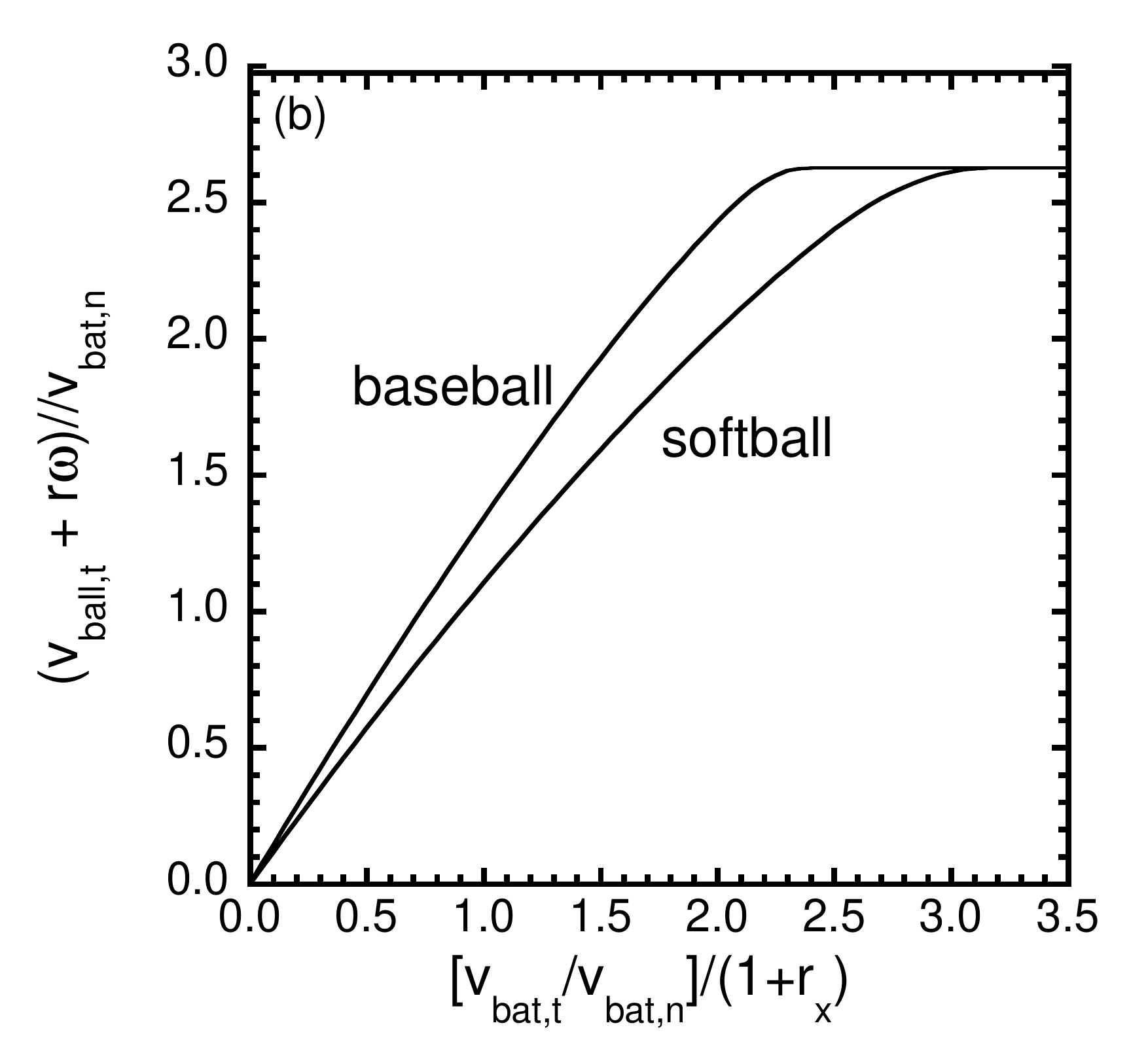}
    \end{minipage}
\caption{\label{fig:bbsbstronge}  Plot of the current data along with a fit using the formalism of Stronge.\cite{Stronge00}  In (a) the solid and open points are for baseball and softball, respectively, and the error bars are estimates based on the fluctuation of the data about a smooth line. In (b), the lines are an extrapolation of the fits to larger impact angles.}
\end{figure}

To apply this model to the present experiment, $\eta^2$ was adjusted to best reproduce the data, resulting in values 4.75 and 6.30 for baseballs and softballs, respectively.  In addition, $e_y$=$\mu$=0.5 was assumed, although the actual results are insensitive to those parameters, at least over the range of $\psi$ of the experiment.  The corresponding tangential vibration frequencies are 0.86$\Omega_n$ and 0.75$\Omega_n$, respectively.  The results are shown in Fig.~\ref{fig:bbsbstronge}, where the LHS of Eq.~\ref{eq:vtf} is plotted against the term in brackets on the RHS.  The local slope is equivalent to the local value of 1+$e_x$.  The calculation does an excellent job accounting for the data and is consistent with the constant value of $e_x$ found in the preceding section.   The present data were all taken at small enough $\psi$ to be consistent with a ``stick-slip'' scenario.  Indeed, the curves are quite linear over a broad range of $\psi$, although they get very nonlinear and eventually flatten after the transition to gross slip.

\begin{figure}[h]
%\begin{center}
\vspace{0.4in}
\singlespace
\includegraphics[angle=0,width=0.7\textwidth]{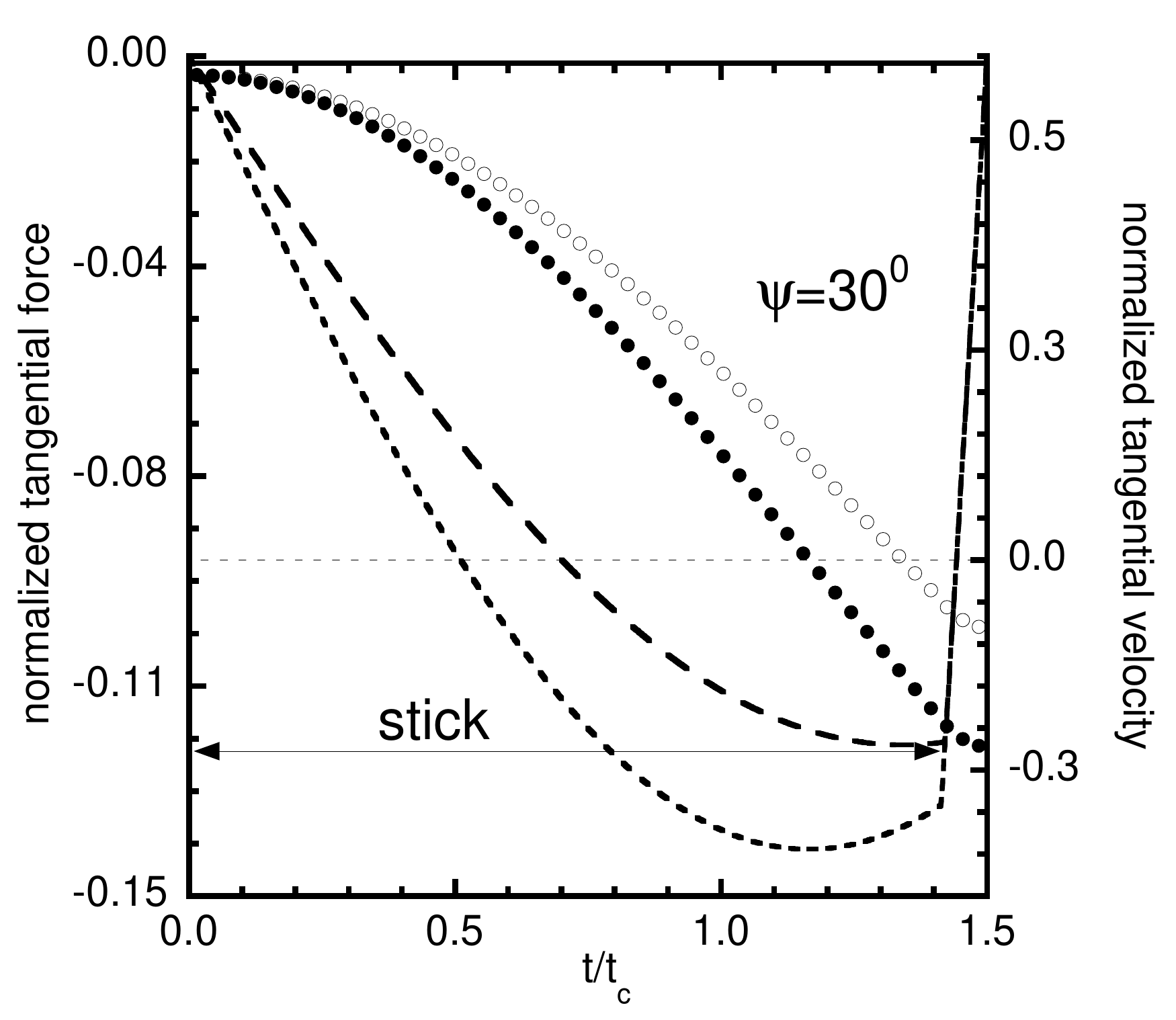}
\caption{\label{fig:stronge}Calculations of the time evolution of the transverse velocity (points, right scale) and force (curves, left scale) for baseballs (closed points and dotted curve) and softballs (open points and dashed curve) in the frame of reference in which the ball is incident at an angle of 30$^\circ$ on a bat at rest.  The velocity is normalized to the initial normal velocity and the force to the peak normal force.  The solid horizontal line indicates the period of initial stick, with a transition to slip at about $t/t_c=1.4$, and the dashed line corresponds to zero transverse velocity.}
%\end{center}
\end{figure}

In Fig.~\ref{fig:stronge}, the transverse velocity and force are plotted for the incident angle $\psi=30^\circ$, in the frame of reference in which the ball is incident on a stationary bat.  For both baseballs and softballs, there is a period of initial stick which persists up to $t\approx 1.4t_c$, followed by a short period of slip until separation at 1.5$t_c$.  During the period of stick, the tangential spring stretches, reaching a maximum at 1.16$t_c$ for baseballs and and 1.33$t_c$ for softballs, then reverses direction just prior to the transition to slip.  After the transition, the transverse force drops rapidly to zero as the spring quickly recovers and separation occurs.  Over the period of stick, the transverse speed is proportional to $\cos{\Omega_t t}$, so that the lower frequency for softballs results in a lower transverse speed at separation and a correspondingly lower value of $e_x$.

As remarked, all the data from the present experiment are consistent with a ``stick-slip'' scenario, whereby the transition to slip occurs very nearly at the end of the collision.  All of the transverse energy dissipated occurs in this final phase and is the result of friction as the contact point slides along the surface.  In effect, a large fraction of the potential energy stored in the transverse spring at the moment of transition gets dissipated in friction.

In the context of this formalism, the primary difference between baseballs and softballs is the difference in tangential compliance, with baseballs being stiffer (less compliant) than softballs.  It remains to be seen whether this property is consistent with the known material properties of these balls.  In that regard, a softball is nearly a uniform sphere, with the bulk of the volume being a rigid polyurethane foam and with a thin leather cover.\cite{Smith2015}  On the other hand, a baseball is a rather complicated object\cite{UML2008} and probably very difficult to model.

\subsection{Implications for batted balls}

\begin{figure}[h]
    \centering
    \begin{minipage}{.5\textwidth}
        \centering
        \includegraphics[width=\textwidth]{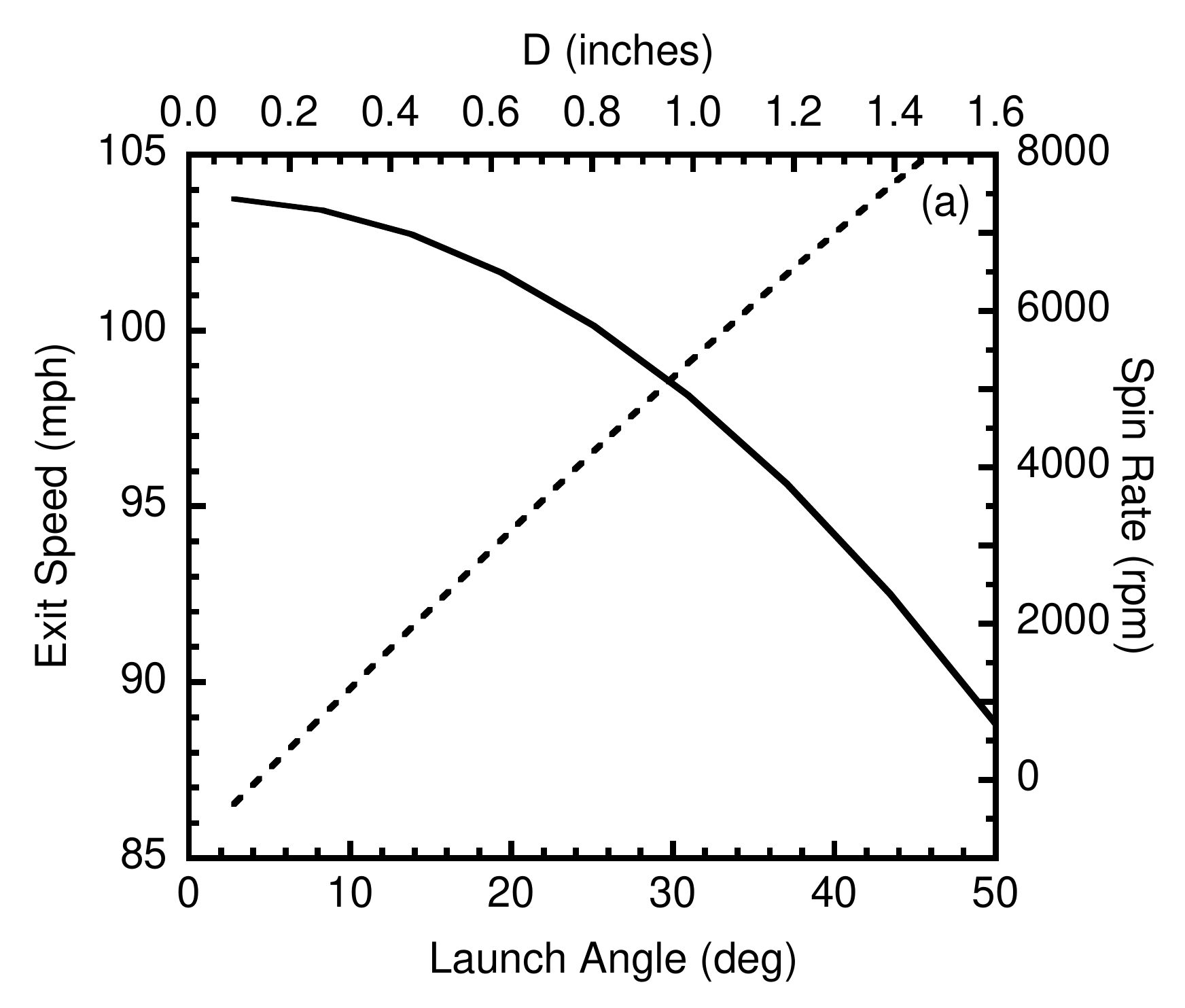}
    \end{minipage}%
    \begin{minipage}{0.5\textwidth}
        \centering
        \includegraphics[width=\textwidth]{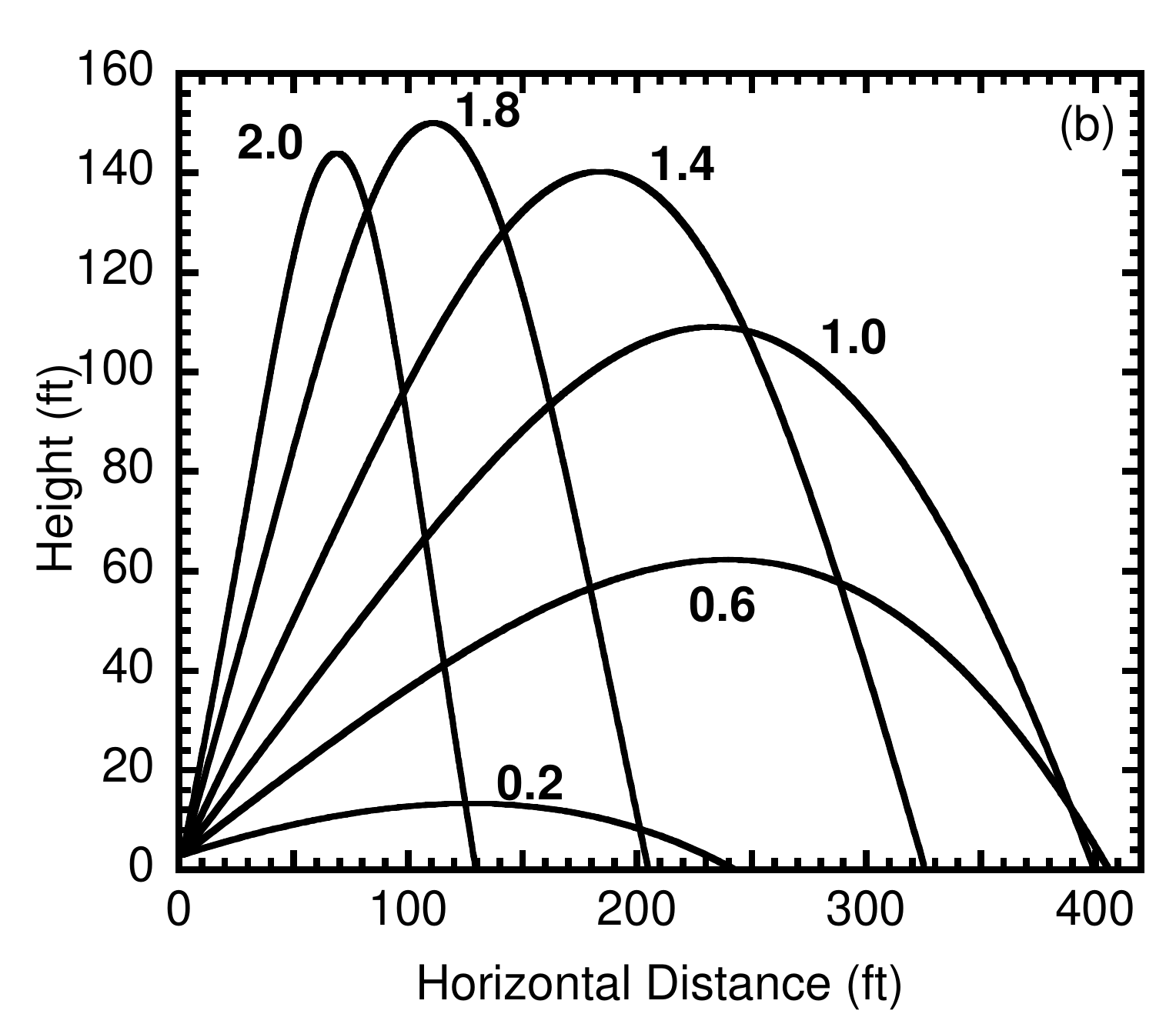}
    \end{minipage}
\caption{\label{fig:CollideCarry}Calculations of the ball-bat collision and the subsequent trajectory.  (a) The exit speed (solid line, left scale) and backspin rate (dashed line, right scale) as a function of both $D$ and launch angle $\theta$.  (b) The trajectory for various values of $D$, indicated next to each curve.}
%\end{center}
\end{figure}

The implication of our results for the spin, speed, launch angle, and fly ball distance of the batted ball are explored next.  This is done only for baseballs, but it would be straightforward to extend the analysis to softballs. An incident pitch was assumed to approach home plate with a speed of 85 mph, a descent angle of 6$^\circ$ degrees, and a backspin rate of 2000 rpm, all values representative of MLB fastballs.\cite{THT1} It was further assumed that the bat was swung at an attack angle of 6$^\circ$ and with a speed of 73 mph at the impact point.   In accordance with the results of the present experiment, $e_x$ was set at $0.4$, where an implicit assumption has been made that the same value of $e_x$ persists at an impact speed of 158 mph as at 77 mph.  Testing that assumption will have to await experiments at higher speed.  The consequences of the resulting collision are explored as a function of the parameter $D$, defined as in Fig.~\ref{fig:geom}, with the results shown in Fig.~\ref{fig:CollideCarry}.  The exit speed and backspin rate as a function of both launch angle and $D$ are shown in the left plot.  As expected, the exit speed peaks at small launch angles ($D\approx 0$), corresponding to a perfect head-on collision, and slowly drops off with increasing launch angle or $D$, in qualitative agreement with data from actual MLB games.\cite{THT3}  The backspin rate is approximately a linear function of $D$.  Note that the rate is negative (i.e., topspin) for $D=0$, a consequence of the incident backspin of the pitch.  The right plot shows the resulting trajectory, calculated from the exit speed, launch angle, and backspin, using drag and lift coefficients previously determined.\cite{Traj} The trajectory evolves from a hard-hit line drive at small $D$, to a 400-ft home run at intermediate $D$, to a lazy fly ball or popup at still larger $D$.  These results conform nicely with observations of trajectories from MLB games.\cite{THT2}

\section{Summary and Conclusions}\label{sec:concl}

Experiments have been made by colliding a swinging bat with a stationary baseball and softball.  The pre-impact velocity of the bat (speed and direction) were measured along with the post-impact velocity and spin of the ball.  The data have been analyzed to determine the tangential coefficient of restitution of each type of ball.  Both baseballs and softballs have spins that are enhanced relative to that expected for a rigid body, suggesting tangential compliance.  In the context of a simple lumped-parameter model of the collision, it is shown that the softballs have a greater tangential compliance than the baseballs.

\begin{acknowledgments}
We thank the group from Rawlings Sporting Goods--Jarden Team Sports, including Biju Mathew,
Wes Lukash, and Chris Flahan, for the use of their facility and help with the data collection.  We also thank Jacob Dahl, Bryant Leung, and Sam Smith from the WSU Sports Science Laboratory for their help with the data collection and reduction.
\end{acknowledgments}

\newpage

\begin{table}[hbt]
 \begin{center}
 \singlespace
 \caption{Mean values of the mass and radii of baseballs used in the present study, with the standard deviation of the quantities in parentheses.  Also shown are the maximum values of $\psi$ and $D$, which are related by Eq.~\ref{eq:D}.}

 \vspace{0.1in}
\begin{tabular}{|c||c|c|c|c|c|}
\hline
ball type        &number    &$m$ (oz) &$r$ (in) &$\psi_{max}$ &$D_{max}$ (in) 		\\ \hline
baseball         &58  &5.04 (0.04)   &1.42 (0.01)&$30^\circ$&1.35\\
softball         &50  &7.09 (0.07)   &1.90 (0.01)&$41^\circ$&2.00 \\

\hline
\hline
\end{tabular}
\end{center}
\label{tab:balls}
\end{table}

\begin{table}[hbt]
 \begin{center}
 \singlespace
 \caption{Properties of bats used in the present study, including the mass $M$, the moments of inertia about two orthogonal axes $I_0$ and $I_z$, and the radius $R$ of the bat at the impact location and its distance $b$ from the center of mass.  Also given are the values of $e_x$ for baseballs and softballs, with standard errors in parentheses.}
 \vspace{0.1in}
\begin{tabular}{|c||c|c|c|c|c|c|c|}
\hline
bat type        &$M$ 	&$I_0$ 	      &$I_z$      &$b$   &$R$ &$e_x$&$e_x$	\\
                &(oz)   &(oz-in$^2$)    &(oz-in$^2$)&(in)  &(in)   &baseball    &softball     \\  \hline
wood            &29.41  &2370   &9.1    &6.8    &1.16   &0.464(14) &0.234(17)\\
metal rough     &30.83  &2758   &18.0   &5.8    &1.28   &0.356(15) &0.130(38)\\
metal smooth    &31.59  &3165   &18.3   &6.6    &1.29   &0.374(15) &0.128(13)\\
softball composite        &28.33  &2867   &11.9   &6.2    &1.13 &--- &0.128(12) \\
\hline
\hline
\end{tabular}
\end{center}
\label{tab:bats}
\end{table}

\newpage

\begin{thebibliography}{}
\bibitem{D} In collision theory, $D$ is referred to as the impact parameter.
\end{thebibliography}
\end{document}